\documentclass[12pt]{iopart}

\usepackage{color}

\usepackage{bm}
\usepackage{amssymb}
\usepackage{graphicx}
\usepackage{array}
\usepackage{tabulary}

\usepackage{etoolbox}
\apptocmd{\sloppy}{\hbadness 10000\relax}{}{}

\newcommand{\tfrac}[2]{{\textstyle{\frac{#1}{#2}}}}
\newcommand{\hf}{\frac{1}{2}}
\newcommand{\thf}{\tfrac{1}{2}}
\newcommand{\thq}{\tfrac{1}{4}}
\newcommand{\thh}{\tfrac{1}{8}}
\newcommand{\bfr}{\mathbf r}
\newcommand{\bfs}{\mathbf s}
\newcommand{\bfk}{\mathbf k}
\newcommand{\bfnab}{\boldsymbol\nabla}
\newcommand{\bfsig}{\boldsymbol\sigma}
\newcommand{\bfj}{\mathbf j}

\newcommand{\bfT}{\mathbf T}

\newcommand{\bbJ}{\mathsf J}

\newcommand{\rhosat}{\rho_\mathrm{sat}}

\newcommand{\overbar}[1]{%
\mkern 1.5mu\overline{\mkern-1.5mu#1\mkern-1.5mu}\mkern 1.5mu}

\newcolumntype{K}[1]{>{\centering\arraybackslash}p{#1}}

\newcommand{\bld}[1]{\boldsymbol{#1}}

\begin{document}

\title{Nonlocal energy density functionals for
pairing and beyond-mean-field calculations}

\author{K. Bennaceur$^{1,2,3}$, A. Idini$^{2,4}$,
        J. Dobaczewski$^{2,3,5,6}$, P.~Dobaczewski$^{7}$,
        M.~Kortelainen$^{2,3}$, and F. Raimondi$^{2,4}$}
\address{$^1$Univ Lyon, Universit\'e Lyon 1, CNRS/IN2P3, IPNL, F-69622 Villeurbanne, France}
\address{$^2$Department of Physics, PO Box 35 (YFL), FI-40014
University of Jyv{\"a}skyl{\"a}, Finland}
\address{$^3$Helsinki Institute of Physics, P.O. Box 64, FI-00014 University of Helsinki, Finland}
\address{$^4$Department of Physics, University of Surrey, Guildford GU2 7XH,
 United Kingdom}
\address{$^5$Department of Physics, University of York, Heslington,
York YO10 5DD, United Kingdom}
\address{$^6$Institute of Theoretical Physics, Faculty of Physics,
University of Warsaw, ul. Pasteura 5, PL-02-093 Warsaw, Poland}
\address{$^7$ul. Obozowa 85 m. 5, PL-01425 Warsaw, Poland}

\begin{abstract}
We propose to use two-body regularized finite-range pseudopotential to generate
nuclear energy density functional (EDF) in both particle-hole and
particle-particle channels, which makes it free from self-interaction
and self-pairing, and also free from singularities when used beyond
mean field. We derive a sequence of pseudopotentials regularized up
to next-to-leading order (NLO) and next-to-next-to-leading order
(N$^2$LO), which fairly well describe infinite-nuclear-matter
properties and finite open-shell paired and/or deformed nuclei.
Since pure two-body pseudopotentials cannot generate sufficiently large
effective mass, the obtained solutions constitute a preliminary
step towards future implementations, which will include, e.g., EDF terms
generated by three-body pseudopotentials.
\end{abstract}

\maketitle

\section{Introduction}

The two most widely used families of non-relativistic nuclear energy
density functionals (EDFs) are based on the
Skyrme~\cite{Skyrme:56,Skyrme:58a} and Gogny~\cite{(Dec80a)}
functional generators. The main difference between them is that the
Skyrme-type generators are built as sums of contact terms with nonlocal
gradient corrections, whereas the Gogny-type ones are built as sums of local
finite-range Gaussians. In Gogny and most Skyrme parametrizations, to
conveniently reproduce properties of homogeneous nuclear matter, a two-body
density-dependent generator is usually used, most often with a fractional
power of the density. Unfortunately, such an approach leads to
non-analytic properties of beyond-mean-field EDF in the complex
plane~\cite{(Dob07)}, compromises symmetry-restoration
procedures~\cite{(Ang01b),(Dob07),Bender:09}, and introduces
self-interaction contributions in the particle-vibration coupling
vertex~\cite{Tarpanov:14}.

Thus, to gain progress in the development of a consistent description
of atomic nucleus, there clearly appears a need to build EDFs that
are free from the above spuriousities. Such functionals can be
obtained by defining their potential parts as Hartree-Fock-Bogolyubov
(HFB) expectation values of genuine pseudopotential operators.
In fact, this was the original idea of Skyrme who first introduced the
name pseudopotential in this context~\cite{(Sky57)}.
Without density-dependent terms and taking all exchange and pairing
terms into account, this gives EDFs for which the Pauli principle is
strictly obeyed and removes all spurious contributions. Following
Refs.~\cite{Raimondi:14,(Dob16)}, also here we call such
pseudopotentials EDF generators.

In Refs.~\cite{(Car08a),(Rai11a),(Dob12b),Raimondi:14}, we have already fully
developed the formalism that uses contact and regularized
higher-order pseudopotentials to generate the most general terms in
the EDF compatible with symmetries. There were several recent
attempts to use such EDFs, like the development of the
Skyrme--inspired family of functionals SLyMR which include 3- and
4-body terms~\cite{(Sad13c)}, our previous parametrization of the
regularized finite-range pseudopotential REG2~\cite{Bennaceur:14}, or
functional VLyB, which implemented higher-order contact
terms~\cite{Davesne:15}. However none of the previous attempts
managed to reproduce both bulk and pairing properties of finite
nuclei, while ensuring at the same time stability of homogeneous
nuclear matter.

The aim of this article is thus to present a further step in the
construction of a predictive pseudopotential-based EDF. We present
an adjustment of a parametrization of the regularized finite-range
higher-order local and density-independent pseudopotential, which
achieves an acceptable qualitative description both in the
particle-hole and particle-particle channels without leading to
infinite-matter instabilities. With the limitation of the current
implementation being only purely two-body and local, a sufficiently
high value of the infinite-matter effective mass could not be
obtained. However, reasonably good results obtained for bulk and
pairing properties of finite nuclei demonstrate proof-of-principle
feasibility of such program.

The article is organized as follows. In section~\ref{sec:regpp} we
briefly recall the form of the pseudopotential and present the
corresponding EDF terms in particle-hole and particle-particle
channels. The numerical implementation of the mean-field equations
in the new HFB spherical solver {\sc{finres}}$_4$ is briefly discussed
in section~\ref{sec:finres}. The strategy used to adjust the
parameters is given in section~\ref{sec:fit}. In
section~\ref{sec:res} we discuss statistical errors of the resulting
parameters and observables, and present selected results for infinite
nuclear matter and finite nuclei. Conclusions are given in
section~\ref{sec:Conclusions}.

\section{Regularized pseudopotential}
\label{sec:regpp}

In this section we recall the Cartesian form of the regularized
pseudopotential as introduced in~\cite{(Dob12b),Raimondi:14}. We
derived the corresponding EDF both in the particle-hole and
particle-particle channels up to N$^3$LO, whereas
below we show them up to NLO.

\subsection{Form factors of the pseudopotential}

The pseudopotential can be regarded as a modified Skyrme interaction
with the $\delta$ form factor replaced by a finite-range regulator,
for which we have chosen a Gaussian form,
\begin{equation}
g_a(\bfr)=\frac{1}{(a\sqrt{\pi})^3}\,\rme^{-\frac{\bfr^2}{a^2}}\,.
\label{eq:g}
\end{equation}
The pseudopotential regularized at order $p$, that is, up to N$^p$LO, depends on
differential operators of the order $n=2p$, and reads
\begin{eqnarray}
{\cal V}^{(n)}_j(\bfr_{1},\bfr_{2};\bfr_{3},\bfr_{4})&=&\left(
 W^{(n)}_j\hat 1_\sigma\hat 1_\tau
+B^{(n)}_j\hat 1_\tau\hat P^\sigma
-H^{(n)}_j\hat 1_\sigma\hat P^\tau
-M^{(n)}_j\hat P^\sigma\hat P^\tau\right) \nonumber \\
&&\times\hat O^{(n)}_j(\bfk_{12},\bfk_{34})
\delta(\bfr_{13})\delta(\bfr_{24})g_a(\bfr_{12})\,.
\label{eq:pp}
\end{eqnarray}
where $\hat 1_\sigma$ and $\hat 1_\tau$ are respectively the identity
operators in spin and isospin space and $\hat P^\sigma$ and $\hat
P^\tau$ the spin and isospin exchange operators. Standard
relative-momentum operators are defined as
$\bfk_{ij}=\tfrac{1}{2i}\left(\bfnab_{i}-\bfnab_{j}\right)$
and relative positions as $\bfr_{ij}=\bfr_{i}-\bfr_{j}$.
Operators $\hat O^{(n)}_j(\bfk_{12},\bfk_{34})$ are scalars built at a
given order $n$ from relative momenta $\bfk_{12}^*$ and $\bfk_{34}$
and index $j$ enumerates such different scalars~\cite{Raimondi:14}.
We note here that coupling constants $W_j^{(n)}$, $B_j^{(n)}$,
$H_j^{(n)}$, and $M_j^{(n)}$ are defined in a different convention
than those of the Gogny effective interaction~\cite{(Dec80a)},
because they include coefficient $\left(a\sqrt{\pi}\right)^{-3}$ introduced
in the form factor in Eq.~(\ref{eq:g}).

At LO, that is for $n=0$, we have $O_0(\bfk_{12},\bfk_{34})=\hat 1$,
which gives a local pseudopotential that reads,
\begin{eqnarray}
\label{eq:LOpot}
{\cal V}^{(0)}_{\mathrm{loc}}(\bfr_{1},\bfr_{2};\bfr_{3},\bfr_{4})&=&\left(
 W^{(0)}_1\hat 1_\sigma\hat 1_\tau
+B^{(0)}_1\hat 1_\tau\hat P^\sigma
-H^{(0)}_1\hat 1_\sigma\hat P^\tau
-M^{(0)}_1\hat P^\sigma\hat P^\tau\right) \nonumber \\
&&\hspace{2cm}\times
\delta(\bfr_{13})\delta(\bfr_{24})g_a(\bfr_{12})\,,
\end{eqnarray}
At a given order higher order $n>0$, the pseudopotential is local if it only depends on
$\bfk_{34}+\bfk_{12}\equiv\bfk_{34}-\bfk_{12}^*$~\cite{(Dob12b),Raimondi:14}.
With the conventions introduced in~\cite{Raimondi:14}, this
corresponds to conditions,
\begin{eqnarray}
\label{eq:local}
\begin{array}{l}
W^{(n)}_2=-W^{(n)}_1,\quad B^{(n)}_2=-B^{(n)}_1, \quad
H^{(n)}_2=-H^{(n)}_1,\quad M^{(n)}_2=-M^{(n)}_1, \\
W^{(n)}_j=B^{(n)}_j=H^{(n)}_j=M^{(n)}_j=0 \quad\mbox{for}\quad j>2.
\end{array}
\end{eqnarray}
The pseudopotential then takes the form
\begin{eqnarray}
\label{eq:NLOpot1}
{\cal V}^{(n)}_{\mathrm{loc}}(\bfr_{1},\bfr_{2};\bfr_{3},\bfr_{4})&=&\left(
 W^{(n)}_1\hat 1_\sigma\hat 1_\tau
+B^{(n)}_1\hat 1_\tau\hat P^\sigma
-H^{(n)}_1\hat 1_\sigma\hat P^\tau
-M^{(n)}_1\hat P^\sigma\hat P^\tau\right) \nonumber \\
&\times&\!\!\left[\hat O^{(n)}_1(\bfk_{12},\bfk_{34})
-\hat O^{(n)}_2(\bfk_{12},\bfk_{34})\right]
\delta(\bfr_{13})\delta(\bfr_{24})g_a(\bfr_{12})\,,
\end{eqnarray}
or
\begin{eqnarray}
\label{eq:NLOpot2}
{\cal V}^{(n)}_{\mathrm{loc}}(\bfr_{1},\bfr_{2};\bfr_{3},\bfr_{4})&=&\left(
 W^{(n)}_1\hat 1_\sigma\hat 1_\tau
+B^{(n)}_1\hat 1_\tau\hat P^\sigma
-H^{(n)}_1\hat 1_\sigma\hat P^\tau
-M^{(n)}_1\hat P^\sigma\hat P^\tau\right) \nonumber \\
&&\times\left(\frac{1}{2}\right)^{n/2}\left[\bfk_{34}+\bfk_{12}\right]^n
\delta(\bfr_{13})\delta(\bfr_{24})g_a(\bfr_{12})\,.
\end{eqnarray}
The fact that the derivative operator commutes with
$\delta(\bfr_{13})\delta(\bfr_{24})$ leads in this case to
the following expression
\begin{eqnarray}
\label{eq:locpot}
&&{\cal V}^{(n)}_{\mathrm{loc}}(\bfr_{1},\bfr_{2};\bfr_{3},\bfr_{4})=\left(W^{(n)}_1\hat 1_\sigma\hat 1_\tau
+B^{(n)}_1\hat 1_\tau\hat P^\sigma
-H^{(n)}_1\hat 1_\sigma\hat P^\tau
-M^{(n)}_1\hat P^\sigma\hat P^\tau\right)  \nonumber \\
&&\hskip 4cm\times\delta(\bfr_{13})\delta(\bfr_{24})
\left(\frac{1}{2}\right)^{n/2}\bfk_{12}^n g_a(\bfr_{12})\,.
\end{eqnarray}
Furthermore, the identity,
\begin{equation}
\left(\frac{1}{2i^2}\right)^{n/2}\bfnab^n g_a(\bfr)=\left(-\frac{1}{2}\right)^p
\Delta^p g_a(\bfr)
=\left(-\frac{1}{a}\frac{\partial}{\partial a}\right)^p g_a(\bfr)\,,
\label{eq:lapg}
\end{equation}
where $p=n/2$, shows that, in the case of a local pseudopotential
(\ref{eq:local}), any contribution to the energy or mean-field
regularized at order $p$ can be obtained from the corresponding
expression at order $p=0$ by iterating $p$ times operator
$-\frac{1}{a}\frac{\partial}{\partial a}$ on it\,.

The notations used in this section are suitable for writing the
pseudopotential and the corresponding functional at any order.
In section~\ref{sec:edf}, we restrict the pseudopotential to
terms up to second only and adopt a lighter notation with
${\cal V}_1^{(0)}\equiv {\cal V}_0$, ${\cal V}_1^{(2)}\equiv {\cal V}_1$,
${\cal V}_2^{(2)}\equiv {\cal V}_2$ and similar conventions for
operators $\hat{O}_j^{(n)}$ and parameters $W_j^{(n)}$,
$B_j^{(n)}$, $H_j^{(n)}$ and $M_j^{(n)}$, so we write,
\begin{eqnarray}
 {\cal V}_k(\bfr_{1},\bfr_{2};\bfr_{3},\bfr_{4})&=&\left(W_k\hat 1_\sigma\hat 1_\tau
+B_k\hat 1_\tau\hat P^\sigma
-H_k\hat 1_\sigma\hat P^\tau
-M_k\hat P^\sigma\hat P^\tau\right) \nonumber \\
&&\times
\hat O_k(\bfk_{12},\bfk_{34})
\delta(\bfr_{13})\delta(\bfr_{24})g_a(\bfr_{12})\,,
\label{eq:pp2}
\end{eqnarray}
with $k=0$, 1, or 2.

%
%

\subsubsection{Nonlocal densities and currents.}

An average value of the energy of a nucleus can be conveniently written
using one-body normal and pairing densities. Their definitions and some
properties~\cite{Perlinska:04} are recalled in this section.

For a given reference state $|\Psi\rangle$, the non-local one-body
(normal) density is defined as
\begin{equation}
\rho(\bfr_1s_1t_1,\bfr_2s_2t_2)=\langle\Psi|
a^\dagger_{\bfr_2s_2t_2}a^{~}_{\bfr_1s_1t_1}|\Psi\rangle\,,
\end{equation}
where the operators $a^\dagger_{\bfr st}$ and $a^{~}_{\bfr st}$
create and annihilate nucleons at position $\bfr$ having spin and
isospin projections $s=\pm\frac{1}{2}$ and $t=\pm\frac{1}{2}$.
The pairing density is defined as
\begin{equation}
\tilde\rho(\bfr_1s_1t_1,\bfr_2s_2t_2)=-2s_2\langle\Psi|
a^{~}_{\bfr_2\mbox{\,-}s_2t_2}a^{~}_{\bfr_1s_1t_1}|\Psi\rangle\,.
\end{equation}
These two densities satisfy the properties
\begin{eqnarray}
&\rho(\bfr_1s_1t_1,\bfr_2s_2t_2)
=\rho^*(\bfr_2s_2t_2,\bfr_1s_1t_1)\,, \\
&{\tilde\rho}(\bfr_1s_1t_1,\bfr_2s_2t_2)
=4s_1s_2\,{\tilde\rho}(\bfr_2\mbox{\,-}s_2t_2,\bfr_1\mbox{\,-}s_1t_1)\,.
\end{eqnarray}

In the present article, we do not consider the possibility to mix
protons and neutrons. This restriction has the consequence that
the normal and pairing densities are diagonal in the neutron-proton space.
In this situation, the normal density can be expanded in
(iso)scalar/(iso)vector parts using identity and Pauli matrices as
\begin{eqnarray}
\rho(\bfr_1s_1t_1,\bfr_2s_2t_2)
&=\thq\,\rho_0(\bfr_1,\bfr_2)\delta_{s_1s_2}\delta_{t_1t_2}
+\thq\,\rho_1(\bfr_1,\bfr_2)\delta_{s_1s_2}\tau^{(3)}_{t_1t_2} \nonumber\\
&+\thq\,\bfs_0(\bfr_1,\bfr_2)\cdot{\bfsig}_{s_1s_2}\delta_{t_1t_2}
+\thq\,\bfs_1(\bfr_1,\bfr_2)\cdot{\bfsig}_{s_1s_2}
\tau^{(3)}_{t_1t_2}\,,
\end{eqnarray}
and the pairing density as
\begin{equation}
{\tilde\rho}(\bfr_1s_1t,\bfr_2s_2t)
=\thf\,\tilde\rho_t(\bfr_1,\bfr_2)\delta_{s_1s_2}
+\thf\,\tilde\bfs_t(\bfr_1,\bfr_2)\cdot{\bfsig}_{s_1s_2}\,.
\end{equation}
In the latter expression, index $t=n$ or $p$ stands for neutrons or protons,
respectively.

Since the pseudopotential may contain derivative terms, we introduce the
following nonlocal densities
\begin{eqnarray}
&\tau_T(\bfr_1,\bfr_2)=\bfnab_1\cdot\bfnab_2\,\rho_T(\bfr_1,\bfr_2)\,, \\
&\bfT_T(\bfr_1,\bfr_2)=\bfnab_1\cdot\bfnab_2\,\bfs_T(\bfr_1,\bfr_2)\,, \\
&\bfj_T(\bfr_1,\bfr_2)=\tfrac{1}{2\rmi}\left(
\bfnab_1-\bfnab_2\right)\rho_T(\bfr_1,\bfr_2)\,, \\
&J_{T\mu\nu}(\bfr_1,\bfr_2)=\tfrac{1}{2\rmi}\left(
\bfnab_1-\bfnab_2\right)_\mu s_{T\nu}(\bfr_1,\bfr_2)
\end{eqnarray}
which are respectively the nonlocal scalar kinetic, pseudo-vector
spin-kinetic, vector current and spin-orbit tensor densities. In these
expressions, index $T=0$ or 1 stands for isoscalar or isovector densities,
respectively and $\mu$ and $\nu$ represent the Cartesian coordinates
in directions $x$, $y$ and $z$.
In a similar manner, we introduce the following non-local pairing densities
\begin{eqnarray}
&\tilde\tau_t(\bfr_1,\bfr_2)
       =\bfnab_1\cdot\bfnab_2\,\tilde\rho_t(\bfr_1,\bfr_2)\,, \\
&\tilde\bfT_t(\bfr_1,\bfr_2)
       =\bfnab_1\cdot\bfnab_2\,\tilde\bfs_t(\bfr_1,\bfr_2)\,, \\
&\tilde{\boldsymbol\jmath}_t(\bfr_1,\bfr_2)=\tfrac{1}{2\rmi}\left(
\bfnab_1-\bfnab_2\right)\tilde\rho_t(\bfr_1,\bfr_2)\,, \\
&\tilde J_{t\mu\nu}(\bfr_1,\bfr_2)=\tfrac{1}{2\rmi}\left(
\bfnab_1-\bfnab_2\right)_\mu\tilde s_{t\nu}(\bfr_1,\bfr_2)\,.
\end{eqnarray}

\subsection{Structure of the nonlocal energy density functional}

\label{sec:edf}

By summing over spin and isospin indices, one obtains
contributions to the energy that come from different terms
of the regularized pseudopotential, as given in equation~(\ref{eq:pp2}).
For given values of $k$, this energy contains the local part,
\begin{eqnarray}
\langle V^L_k\rangle&=&\sum_{T=0,1}\int
\rmd^3r_1\,\rmd^3r_2\,\rmd^3r_3\,\rmd^3r_4\left[\hat O_k(\bfk_{12},\bfk_{34})
\delta(\bfr_{13})\delta(\bfr_{24})g_a(\bfr_{12})\right] \nonumber \\
&&\times\Bigl[
 A_k^{\rho_T}\rho_T(\bfr_3,\bfr_1)\rho_T(\bfr_4,\bfr_2)
+A_k^{\bfs_T}\bfs_T(\bfr_3,\bfr_1)\cdot\bfs_T(\bfr_4,\bfr_2)
\Bigr]\,,
\label{eq:vjl}
\end{eqnarray}
the nonlocal part,
\begin{eqnarray}
\langle V^N_k\rangle&=&\sum_{T=0,1}\int
\rmd^3r_1\,\rmd^3r_2\,\rmd^3r_3\,\rmd^3r_4\left[\hat O_k(\bfk_{12},\bfk_{34})
\delta(\bfr_{13})\delta(\bfr_{24})g_a(\bfr_{12})\right] \nonumber \\
&&\times\Bigl[
 B_k^{\rho_T}\rho_T(\bfr_4,\bfr_1)     \rho_T(\bfr_3,\bfr_2)
+B_k^{\bfs_T}\bfs_T(\bfr_4,\bfr_1)\cdot\bfs_T(\bfr_3,\bfr_2)
\Bigr]\,,
\label{eq:vjn}
\end{eqnarray}
and the pairing part
\begin{eqnarray}
\langle V^P_k\rangle&=&\sum_{t=n,p}\int
\rmd^3r_1\,\rmd^3r_2\,\rmd^3r_3\,\rmd^3r_4\left[\hat O_k(\bfk_{12},\bfk_{34})
\delta(\bfr_{13})\delta(\bfr_{24})g_a(\bfr_{12})\right] \nonumber \\
&&\times\Bigl[
 C_k^{\tilde\rho}\,\tilde\rho_t^*(\bfr_1,\bfr_2)     \tilde\rho_t(\bfr_3,\bfr_4)
+C_k^{\tilde\bfs}\,\tilde\bfs_t^*(\bfr_1,\bfr_2)\cdot\tilde\bfs_t(\bfr_3,\bfr_4)
\Bigr]\,.
\label{eq:vjp}
\end{eqnarray}
Expressions for coupling constants $A$, $B$, and $C$ that appear in Eqs.~(\ref{eq:vjl}),
(\ref{eq:vjn}), and~(\ref{eq:vjp}) read
\begin{eqnarray}
A_k^{\rho_0}&= \thf\,W_k+\thq\,B_k-\thq\,H_k-\thh\,M_k\,,\\
A_k^{\bfs_0}&=           \thq\,B_k          -\thh\,M_k\,,\\
B_k^{\rho_0}&=-\thh\,W_k-\thq\,B_k+\thq\,H_k+\thf\,M_k\,,\\
B_k^{\bfs_0}&=-\thh\,W_k          +\thq\,H_k          \,,\\
A_k^{\rho_1}&=                    -\thq\,H_k-\thh\,M_k\,,\\
A_k^{\bfs_1}&=                              -\thh\,M_k\,,\\
B_k^{\rho_1}&=-\thh\,W_k-\thq\,B_k                    \,,\\
B_k^{\bfs_1}&=-\thh\,W_k                              \,,\\
C_k^{\tilde\rho}&= \thq\,W_k-\thq\,B_k-\thq\,H_k+\thq\,M_k\,,\\
C_k^{\tilde\bfs}&= \thq\,W_k+\thq\,B_k-\thq\,H_k-\thq\,M_k\,.
\end{eqnarray}
%

%
%

\subsubsection{Leading-order term of the pseudopotential.}

The leading-order pseudopotential is modeled by a simple
Gaussian form factor and does not contain derivative terms. In
this case, operator $\hat O_0$ simply reads
\begin{equation}
\hat O_0(\bfk_{12},\bfk_{34})=\hat 1\,,
\end{equation}
and Eqs.~(\ref{eq:vjl}), (\ref{eq:vjn}), and~(\ref{eq:vjp}) become
\begin{eqnarray}
\langle V^L_0\rangle&=&\sum_{T=0,1}\int
\rmd^3r_1\,\rmd^3r_2\,g_a(\bfr_{12})
\Bigl[
A_0^{\rho_T}\rho_T(\bfr_1)     \rho_T(\bfr_2) \nonumber \\
&&\hspace{4cm}
+A_0^{\bfs_T}\bfs_T(\bfr_1)\cdot\bfs_T(\bfr_2)
\Bigr]\,,  \label{eq:energy0l} \\
\langle V^N_0\rangle&=&\sum_{T=0,1}\int
\rmd^3r_1\,\rmd^3r_2\,g_a(\bfr_{12})
\Bigl[
B_0^{\rho_T}\rho_T(\bfr_2,\bfr_1)     \rho_T(\bfr_1,\bfr_2) \nonumber  \\
&&\hspace{4cm}+B_0^{\bfs_T}\bfs_T(\bfr_2,\bfr_1)\cdot\bfs_T(\bfr_1,\bfr_2)
\Bigr]\,,  \label{eq:energy0n} \\
\langle V^P_0\rangle&=&\sum_{t=n,p}\int
\rmd^3r_1\,\rmd^3r_2\,
g_a(\bfr_{12})
\Bigl[
C_0^{\tilde\rho}\,\tilde\rho_t^*(\bfr_1,\bfr_2)     \tilde\rho_t(\bfr_1,\bfr_2) \nonumber \\
&&\hspace{4cm}+C_0^{\tilde\bfs}\,\tilde\bfs_t^*(\bfr_1,\bfr_2)\cdot\tilde\bfs_t(\bfr_1,\bfr_2)
\Bigr]\,.  \label{eq:energy0p}
\end{eqnarray}

\subsubsection{Next-to-leading order term of the pseudopotential}
\label{sec:intnlo}

At NLO, the two derivative operators are
\begin{equation}
\hat O_{1}(\bfk_{12},\bfk_{34})
=\thf\left(\bfk_{12}^{*2}+\bfk_{34}^2\right)\,,
\end{equation}
and
\begin{equation}
\hat O_{2}(\bfk_{12},\bfk_{34})
=\bfk_{12}^*\cdot\bfk_{34}\,,
\end{equation}
The contribution from the first one to the EDF is given by
\begin{eqnarray}
\langle V_1^L\rangle&=&\hf\sum_{T=0,1}
\int\rmd^3r_1\,\rmd^3r_2\,
g_a(\bfr_{12})  \nonumber\\
&\times&\biggl\{A_1^{\rho_T}\Bigl[\tau_T(\bfr_1)\rho_T(\bfr_2)
-\tfrac{3}{4}\,\rho_T(\bfr_1)\Delta_2\rho_T(\bfr_2)
-\bfj_T(\bfr_1)\cdot\bfj_T(\bfr_2) \Bigr] \nonumber\\
&+&A_1^{\bfs_T}\Bigl[\bfT_T(\bfr_1)\cdot\bfs_T(\bfr_2)
-\tfrac{3}{4}\,\bfs_T(\bfr_2)\cdot\Delta_1\bfs_T(\bfr_1) \nonumber \\
&-&\sum_{\mu\nu}J_{T\mu\nu}(\bfr_1)J_{T\mu\nu}(\bfr_2) \Bigr] \biggr\}\,,
\end{eqnarray}
\begin{eqnarray}
\langle V_1^N\rangle&=&-\hf\!\sum_{T=0,1}\!
\int\!\rmd^3r_1\,\rmd^3r_2  \nonumber\\
&\times&\biggl\{ \frac{1}{2}\Delta g_a(\bfr_{12})
\left[ B_1^{\rho_T}\rho_T(\bfr_2,\bfr_1)\rho_T(\bfr_1,\bfr_2)
+B_1^{\bfs_T}\bfs_T(\bfr_2,\bfr_1)\cdot\bfs_T(\bfr_1,\bfr_2)
\right]\nonumber \\
&+&B_1^{\rho_T}g_a(\bfr_{12})\Bigl[
\rho_T(\bfr_2,\bfr_1)\tau_T(\bfr_1,\bfr_2)
+
\rho_T(\bfr_2,\bfr_1)\Delta_1\rho_T(\bfr_1,\bfr_2) \Bigr] \nonumber\\
&+&B_1^{\bfs_T}g_a(\bfr_{12})\Bigl[
\bfs_T(\bfr_2,\bfr_1)\cdot\bfT_T(\bfr_1,\bfr_2)
+\bfs_T(\bfr_2,\bfr_1)\cdot\Delta_1\bfs_T(\bfr_1,\bfr_2) \Bigr]
 \biggr\}\,,
\end{eqnarray}
and
\begin{eqnarray}
\langle V_1^P\rangle&=&\hf\sum_{t=n,p}
\int\rmd^3r_1\,\rmd^3r_2\,
g_a(\bfr_{12})  \nonumber\\
&\times&\biggl\{C_1^{\tilde\rho}\Bigl[
\tilde\rho_t^*(\bfr_1,\bfr_2)\tilde\tau_t(\bfr_1,\bfr_2)
-\tilde\rho_t^*(\bfr_1,\bfr_2)\Delta_1\tilde\rho_t(\bfr_1,\bfr_2) \Bigr]
  \nonumber\\
&+&C_1^{\tilde\bfs}\Bigl[
\tilde\bfs_t^*(\bfr_1,\bfr_2)\cdot\tilde\bfT_t(\bfr_1,\bfr_2)
-\tilde\bfs_t^*(\bfr_1,\bfr_2)\cdot\Delta_1\tilde\bfs_t(\bfr_1,\bfr_2) \Bigr]
 \biggr\}\,,
\end{eqnarray}
and in turn the contribution from the second one is
\begin{eqnarray}
\langle V_2^L\rangle&=&\hf\sum_{T=0,1}
\int\rmd^3r_1\,\rmd^3r_2\,
g_a(\bfr_{12})   \nonumber\\
&\times&\biggl\{A_2^{\rho_T}\Bigl[\rho_T(\bfr_1)\tau_T(\bfr_2)
+\tfrac{1}{4}\,\rho_T(\bfr_1)\Delta_2\rho_T(\bfr_2)
-\bfj_T(\bfr_1)\cdot\bfj_T(\bfr_2) \Bigr] \nonumber\\
&+&A_2^{\bfs_T}\Bigl[\bfs_T(\bfr_1)\cdot\bfT_T(\bfr_2)
+\tfrac{1}{4}\,\bfs_T(\bfr_2)\cdot\Delta_1\bfs_T(\bfr_1) \nonumber \\
&-&\sum_{\mu\nu} J_{T\mu\nu}(\bfr_1)J_{T\mu\nu}(\bfr_2) \Bigr] \biggr\}\,,
\end{eqnarray}
\begin{eqnarray}
\langle V_2^N\rangle&=&\hf\sum_{T=0,1}
\int\rmd^3r_1\,\rmd^3r_2\,  \nonumber\\
&\times&\biggl\{
\frac{1}{2}\Delta g_a(\bfr_{12})\left[
 B_2^{\rho_T}\rho_T(\bfr_2,\bfr_1)\rho_T(\bfr_1,\bfr_2)
+B_2^{\bfs_T}\bfs_T(\bfr_2,\bfr_1)\cdot\bfs_T(\bfr_1,\bfr_2)
\right] \nonumber \\
&-&
B_2^{\rho_T}g_a(\bfr_{12})\Bigl[
\rho_T(\bfr_2,\bfr_1)\Delta_{1}\rho_T(\bfr_1,\bfr_2)
+\rho_T(\bfr_2,\bfr_1)\tau_T(\bfr_1,\bfr_2) \Bigr] \nonumber\\
&-&
B_2^{\bfs_T}g_a(\bfr_{12})\Bigl[
\bfs_T(\bfr_2,\bfr_1)\cdot\Delta_1\bfs_T(\bfr_1,\bfr_2)
+\bfs_T(\bfr_2,\bfr_1)\cdot\bfT_T(\bfr_1,\bfr_2) \Bigr]
 \biggr\}\,,
\end{eqnarray}
and
\begin{eqnarray}
\langle V_2^P\rangle&=&\frac{1}{2}\sum_{t=n,p}
\int\rmd^3r_1\,\rmd^3r_2 \nonumber \\
&\times&\biggl\{\Delta g_a(\bfr_{12})
 \left[C_2^{\tilde\rho}
\,\tilde\rho_t^*(\bfr_1,\bfr_2)\tilde\rho_t(\bfr_1,\bfr_2)
+C_2^{\tilde\bfs}
\,\tilde\bfs_t^*(\bfr_1,\bfr_2)\cdot\tilde\bfs_t(\bfr_1,\bfr_2)
 \right] \nonumber \\
&+&
C_2^{\tilde\rho}g_a(\bfr_{12})\Bigl[
\tilde\rho_t^*(\bfr_1,\bfr_2)\tilde\tau_t(\bfr_1,\bfr_2)
-\tilde\rho_t^*(\bfr_1,\bfr_2)\Delta_1\tilde\rho_t(\bfr_1,\bfr_2) \Bigr]
 \nonumber \\
&+&
C_2^{\tilde\bfs}g_a(\bfr_{12})\Bigl[
\tilde\bfs_t^*(\bfr_1,\bfr_2)\cdot\tilde\bfT_t(\bfr_1,\bfr_2)
-\tilde\bfs_t^*(\bfr_1,\bfr_2)\cdot\Delta_1\tilde\bfs_t(\bfr_1,\bfr_2) \Bigr]
\biggr\}\,.
\end{eqnarray}

%
%

\subsubsection{Sum of pseudopotential terms at NLO.}

The sum of the terms at NLO leads to the expression where
contributions from the local and nonlocal parts of the pseudopotential
can be explicitly separated. This leads to a form where we can
take advantage of property~(\ref{eq:lapg}), and which is therefore simpler
to implement numerically.

Before we write down the sum of terms given by the two NLO pseudopotential terms, it is
convenient to introduce the following notation for half sum and
half difference of the NLO coupling constants, for example
\begin{eqnarray}
A^{\rho_0}_{12}&=\hf\left(A^{\rho_0}_1+A^{\rho_0}_2\right)\,,\\
A^{\rho_0}_{\overbar{12}}&=\hf\left(A^{\rho_0}_1-A^{\rho_0}_2\right)\,.
\end{eqnarray}
This notation allows us to separate terms of the functional that
come from the local part of the pseudopotential (parameterized by terms
with indices $\overbar{12}$) from those that come from the
nonlocal part of pseudopotential (parameterized by terms
with indices $12$).
The sum of NLO terms shown in section~\ref{sec:intnlo} then reads
\begin{eqnarray}
&&\langle V_1^L\rangle+\langle V_2^L\rangle=\sum_{T=0,1}
\int\rmd^3r_1\,\rmd^3r_2  \nonumber\\
&\times&\biggl\{A_{12}^{\rho_T}g_a(\bfr_{12})
\Bigl[\tau_T(\bfr_1)\rho_T(\bfr_2)
-\tfrac{1}{4}\,\rho_T(\bfr_1)\Delta_2\rho_T(\bfr_2)
-\bfj_T(\bfr_1)\cdot\bfj_T(\bfr_2) \Bigr] \nonumber\\
&+&A_{12}^{\bfs_T}g_a(\bfr_{12})
\Bigl[\bfT_T(\bfr_1)\cdot\bfs_T(\bfr_2)
-\tfrac{1}{4}\,\bfs_T(\bfr_2)\cdot\Delta_1\bfs_T(\bfr_1)
-\bbJ_T(\bfr_1)\cdot\bbJ_T(\bfr_2) \Bigr] \nonumber\\
&-&\hf\Delta g_a(\bfr_{12})
\left[A_{\overbar{12}}^{\rho_T}\rho_T(\bfr_1)\rho_T(\bfr_2)
+A_{\overbar{12}}^{\bfs_T}
\bfs_T(\bfr_1)\cdot\bfs_T(\bfr_2)\right]
 \biggr\}\,, \label{eq:energy12l}
\end{eqnarray}
\begin{eqnarray}
&&\langle V_1^N\rangle+\langle V_2^N\rangle=-\sum_{T=0,1}
\int\rmd^3r_1\,\rmd^3r_2  \nonumber\\
&\times&\biggl\{B_{12}^{\rho_T}g_a(\bfr_{12})\Bigl[
\rho_T(\bfr_2,\bfr_1)\Delta_1\rho_T(\bfr_1,\bfr_2)
+\rho_T(\bfr_2,\bfr_1)\tau_T(\bfr_1,\bfr_2) \Bigr] \nonumber\\
&+&B_{12}^{\bfs_T}g_a(\bfr_{12})\Bigl[
\bfs_T(\bfr_2,\bfr_1)\cdot\Delta_1\bfs_T(\bfr_1,\bfr_2)
+\bfs_T(\bfr_2,\bfr_1)\cdot\bfT_T(\bfr_1,\bfr_2) \Bigr] \nonumber \\
&+&\hf\,\Delta g_a(\bfr_{12})\left[B_{\overbar{12}}^{\rho_T}
\rho_T(\bfr_2,\bfr_1)\rho_T(\bfr_1,\bfr_2)
+B_{\overbar{12}}^{\bfs_T}
\bfs_T(\bfr_2,\bfr_1)\cdot\bfs_T(\bfr_1,\bfr_2)\right]
 \biggr\} \label{eq:energy12n}
\end{eqnarray}
and
\begin{eqnarray}
&&\langle V_1^P\rangle+\langle V_2^P\rangle=\sum_{t=n,p}
\int\rmd^3r_1\,\rmd^3r_2 \nonumber\\
&\times&\biggl\{
C_{12}^{\tilde\rho}\,g_a(\bfr_{12})\Bigl[
\tilde\rho_t^*(\bfr_1,\bfr_2)\tilde\tau_t(\bfr_1,\bfr_2)
-\tilde\rho_t^*(\bfr_1,\bfr_2)\Delta_1\tilde\rho_t(\bfr_1,\bfr_2) \Bigr]
 \nonumber\\
&+&C_{12}^{\tilde\bfs}\,g_a(\bfr_{12})\Bigl[
\tilde\bfs_t^*(\bfr_1,\bfr_2)\cdot\tilde\bfT_t(\bfr_1,\bfr_2)
-\tilde\bfs_t^*(\bfr_1,\bfr_2)\cdot\Delta_1\tilde\bfs_t(\bfr_1,\bfr_2) \Bigr]
 \nonumber\\
&-&\frac{1}{2}\,\Delta g_a(\bfr_{12})\left[C_{\overbar{12}}^{\rho}
\rho_t^*(\bfr_1,\bfr_2)\rho_t(\bfr_1,\bfr_2)
+C_{\overbar{12}}^{\bfs}
\bfs_t^*(\bfr_1,\bfr_2)\cdot\bfs_t(\bfr_1,\bfr_2)\right]
 \biggr\}\,. \label{eq:energy12p}
\end{eqnarray}
One can easily check that when the pseudopotential is reduced to its local part,
three Eqs.~(\ref{eq:energy12l}),
(\ref{eq:energy12n}), and (\ref{eq:energy12p}) can respectively be obtained
from Eqs.~(\ref{eq:energy0l}), (\ref{eq:energy0n}), and (\ref{eq:energy0p})
by using property~(\ref{eq:lapg}).

%
%

\section{Numerical implementations}

\label{sec:finres}

The use of a finite-range pseudopotential makes the mean-field equations a set of coupled non-linear
integro-differential equations, as much as implementing Coulomb interaction exactly and two-body centre-of-mass
correction. In this study, we solve this set of
equations in spherical symmetry using a newly developed code
{\sc{finres}}$_4$ (Finite-Range Self-consistent Spherical
Space-coordinate Solver)~\cite{[Ben17a]}, which is based on the method
proposed by Hooverman~\cite{HOOVERMAN1972155}. With this method, the
differential and integral operators take a form of square matrices,
whereas local fields are simply represented by diagonal ones.

Specifically, densities, fields, and wave functions are discretized in a
spherical box of radius $R$ on a mesh with spacing $\delta r$ starting at
$r=\delta r/2$. The numerical parameters are then $R$, $\delta r$ and
$\ell_\mathrm{max}$. The latter parameter corresponds to the maximum value
of the orbital angular momentum in the partial-wave expansion of one-body
nonlocal densities. The boundary values of wave functions are fixed by
the finite difference formulae used to calculate their derivatives near
$r=R$. In this work, we have chosen to have vanishing wave functions at
$r=R$ in all partial waves.

For deformed nuclei, we used code {\sc{hfodd}}
(v2.78g)~\cite{[Dob17a]}, that is a new version based on previous
releases~\cite{Schunck2012166,(Sch16b)}, in which we implemented
self-consistent solutions for finite-range higher-order
pseudopotentials. Calculations were performed using Cartesian
deformed harmonic-oscillator basis with states included up to
$N_0=16$ shells.

\section{Adjustments of coupling constants}
\label{sec:fit}

In the present implementation, the regularized finite-range local
pseudopotential was supplemented by the Coulomb term and standard zero-range
spin-orbit term~\cite{[Bel56],Skyrme:58a}, and, as discussed below, by a
zero-range two-body term that acts only in the particle-hole channel.
The spin-orbit term was not included in the pairing channel. Effects
of the spurious centre-of-mass motion were approximately removed by the
standard technique consisting in subtracting from the functional
(before variation and without the one-body approximation) the average
value of the momentum squared divided by twice total
mass~\cite{Bender:00}. Adjustments of coupling constants were
performed in spherical symmetry. For simplicity, small
contributions to pairing terms that come from the spin-orbit
interaction were neglected. In addition, for deformed nuclei
calculations were performed with contributions to pairing terms that
come from the Coulomb term also neglected. These two latter
restrictions will be released in future implementations.

Coupling constants of the regularized pseudopotential were determined
by building and minimizing a penalty function whose content is
discussed below. For a given value of the regularization range $a$,
the regularized finite-range local pseudopotential (\ref{eq:local}),
depends on 8 independent parameters at NLO and on 12 at N$^2$LO. The
strength of the spin-orbit interaction makes one additional parameter
$W_\mathrm{SO}$.

After performing several preliminary adjustments, we noted that
pairing fields were strongly peaked at the nuclear surface.
This feature has two unwanted consequences. Firstly, pairing energies and average
gaps were becoming unreasonably strong in neutron rich
isotopes, where a neutron skin develops. Secondly, proton pairing gaps
were much too weak compared to typical expected values.
This was due to the fact that the Coulomb barrier shifts the proton
density from the surface to the nuclear interior.

As a solution to these problems, we considered adding to the
functional a term that makes pairing stronger in the volume without
increasing its strength at the surface. Equivalently, the one
which balances the finite-range pseudopotential, so that it can be
stronger in the pairing channel, without letting the particle-hole channel
becoming too attractive. This can be achieved by adding a zero-range
term of the standard Skyrme type,
\begin{equation}
\label{eq:delta}
{\cal V}_\delta(\bfr_{1},\bfr_{2};\bfr_{3},\bfr_{4})=
t_0\left(1+x_0\hat P^\sigma\right) \delta(\bfr_{13})\delta(\bfr_{24})\delta(\bfr_{12})\,,
\end{equation}
with $x_0=1$. This zero-range term indeed allows us to de-correlate
the behaviour of the LO terms in the particle-hole and $T=1$
particle-particle channels. Since this term does not act in the
pairing channel, no pairing cut-off is needed.
However, when used in beyond-mean-field applications, this term may
still lead to ultra-violet divergence. In principle, it would be
very easy to avoid this by regularizing this term with a short-range
regulator. This route may be exploited in future developments.

A series of tests
performed at NLO showed that for regularization ranges between
$a=1.1$ and 1.3\,fm, the value of
$t_0=1000~\mathrm{MeV}\,\mathrm{fm}^3$ leads to a pairing field which
is not too strongly peaked at the surface. In order to limit the
number of free parameters, this value for $t_0$ was fixed and kept
constant for all pseudopotentials at NLO and N$^2$LO.
We note that by adopting a fixed value of parameter $t_0$, we removed
its influence on the error budget discussed in section~\ref{error}.

\subsection{The penalty function}
\label{sec:penalty}

The penalty function~\cite{0954-3899-41-7-074001} depends
on the vector of parameters of the model $\bld{p}$,
\begin{equation}\label{chi2}
\chi^2(\bld{p})
=
\sum_{i=1}^{N_d}
\frac{\left(\mathcal{O}_i(\bld{p})
      -
      \mathcal{O}^\mathrm{target}_i\right)^2}
     {\Delta\mathcal{O}_i^2} ,
\end{equation}
and measures quadratic
deviation between calculated $\mathcal{O}_i(\bld{p})$ and target
$\mathcal{O}^\mathrm{target}_i$ values for a set of observables
or pseudo-observables with given adopted uncertainties
$\Delta\mathcal{O}_i$. In our implementation, we built the
penalty function as a sum of six different components,
\begin{equation}\label{chi3}
\chi^2
=
 \chi^2_\mathrm{inm}
+\chi^2_\mathrm{pol}
+\chi^2_\mathrm{BE}
+\chi^2_\mathrm{rad}
+\chi^2_\mathrm{gap}
+\chi^2_{\rho_1},
\end{equation}
which are defined as follows:

\begin{itemize}
\item We constrained the following properties of the saturation point
of the infinite nuclear matter, see~\cite{Bennaceur:14} for
definitions, with their adopted uncertainties: saturation density
$\rhosat=0.160\pm0.0005~\mathrm{fm}^{-3}$, binding energy per nucleon
$E/A=-16.00\pm 0.05$\,MeV, incompressibility modulus $K_\infty=230\pm
1$\,MeV, symmetry energy $J=32.0\pm 0.1$\,MeV and its slope
$L=50\pm10$\,MeV. The sum of contributions from these properties
to the penalty function is denoted $\chi^2_\mathrm{inm}$.

\item One value for the energy per nucleon in polarized matter $B_\uparrow(0.16)
=35\pm1$\,MeV. This value does not correspond to the result from a
microscopic calculation and is only considered to prevent the collapse
of polarized matter near the saturation point. We denote the contribution
from this constraint to the penalty function as $\chi^2_\mathrm{pol}$.

\item Binding energies and proton radii of several doubly magic and semi-magic nuclei
as summarized in Table~\ref{tab:one}. Contributions to the penalty function
from theses quantities are denoted $\chi^2_\mathrm{BE}$ and
$\chi^2_\mathrm{rad}$, respectively.
\item
The zero-range term (\ref{eq:delta}) turned out to be useful but not
sufficient to guarantee that the pairing field would not be too
strong at the nuclear surface. For that purpose, we used an
additional scheme when constraining average pairing gaps. Since all
our densities are expended in partial waves up to a given value
$\ell_\mathrm{max}$, a pairing field strong at the surface is
expected to give a significant change of the results if the maximum
value of $\ell$ is changed from $\ell_\mathrm{max}=\ell_0$ to
$\ell_\mathrm{max}=\ell_0+2$. Reciprocally, an average neutron
pairing gap in a given nucleus, which is constrained to give
approximately the same value for calculations with
$\ell_\mathrm{max}=\ell_0$ and $\ell_\mathrm{max}=\ell_0+2$ can
prevent the pairing field from being too strong near the surface.

In practice, we constrained the average neutron pairing gap
$\langle\Delta_n\rangle$ in $^{120}$Sn calculated at
$\ell_\mathrm{max}=9$ and $\ell_\mathrm{max}=11$. Pseudopotentials
considered here lead to a low effective mass and thus a low density
of states. In order to avoid too frequent collapses of pairing
correlations in nuclei with subshells closure or with the opening of
gaps for deformed nuclei, we decided to largely overshoot the
value of the average pairing gap compared to what can be extracted
from experimental mass staggering, and we used the target value of
$\langle\Delta_n\rangle=2.8$\,MeV for the two values of
$\ell_\mathrm{max}$ with a small uncertainty of $0.002$\,MeV. This
ensures that the average pairing gap is almost the same for the two
truncations and therefore the pairing field does not significantly
change when more partial waves are added. The contribution from these
constraints to the penalty function is denoted $\chi^2_\mathrm{gap}$.

\item
We have observed that the fit of the parameters can easily drive the
pseudopotential into regions of the parameter space that lead to finite-size
instabilities similar to those already identified for Skyrme
functionals~\cite{PhysRevC.88.064323}. For this latter type of functionals, a
tool based on the linear response theory was developed to characterize and
avoid these finite-size instabilities. Such a tool does not yet exist
for the pseudopotentials considered here, so we had to rely on an empirical
criterion. We noticed that before the parameters of a pseudopotential
get close to a region where finite-size isovector instabilities develop,
strong oscillations can be seen in the isovector density $\rho_1(\bfr)$ of heavy
nuclei. Specifically, those oscillations lead to a decrease of the density of
neutrons and increase of density of the protons at the centre of $^{208}$Pb
so that $\rho_1(0)<0$. To avoid the appearance of the finite-size
isovector instabilities, we introduced a constraint from the central
isovector density $\rho_1(0)$ in $^{208}$Pb to enforce $\rho_1(0)>0$, that is,
we have calculated the quantity,
\begin{equation}
C=\exp\left[-\frac{\rho_1(0)}{\alpha}\right]\,,
\end{equation}
which contributes to the penalty function as
\begin{equation}
\chi^2_{\rho_1}=\left(\frac{C-C^\mathrm{target}}{\Delta C}\right)^2,
\end{equation}
with the targeted value of $C^\mathrm{target}=0$ and the uncertainty
$\Delta C=1$. Parameter $\alpha$ was here empirically set to
$0.006~\mathrm{fm}^{-3}$.
\end{itemize}

We note that the adopted structure of the penalty function mixes real
experimental data and metadata, the latter certainly introducing some
poorly controlled bias to the fit. Unfortunately, the use of metadata
seems to be unavoidable, in the sense that the real experimental data
do not alone constrain the model parameters sufficiently. As a
result, without constraints on metadata, the fits easily drift
towards clearly unphysical regions of the parameter space, and thus
become useless. These aspects must become a centrepiece of future
investigations in this domain.

\begin{table}[htbp]
\begin{center}
\begin{tabular}{|l|r|r|r|r|}
\hline
Nucleus & $E_\mathrm{exp}~[\mathrm{MeV}]$ &
          $\Delta E_\mathrm{exp}~[\mathrm{MeV}]$ &
          $r_\mathrm{p}~[\mathrm{fm}]$ &
          $\Delta r_\mathrm{p}~[\mathrm{fm}]$ \\
\hline
$^{40}$Ca   &  -342.034 & 1.000 & 3.382 & 0.020 \\
$^{48}$Ca   &  -415.981 & 1.000 & 3.390 & 0.020 \\
$^{56}$Ni   &  -483.954 & 1.000 & 3.661 & 0.020 \\
$^{78}$Ni   &  -641.743 & 2.000 &       &       \\
$^{100}$Sn  &  -824.775 & 1.000 &       &       \\
$^{120}$Sn  & -1020.375 & 3.000 &       &       \\
$^{132}$Sn  & -1102.680 & 1.000 &       &       \\
$^{208}$Pb  & -1635.893 & 1.000 & 5.450 & 0.020 \\
\hline
\end{tabular}
\caption{Binding energies and proton radii used in the partial
penalty functions $\chi^2_\mathrm{BE}$ and $\chi^2_\mathrm{rad}$,
respectively. The binding energy of $^{78}$Ni and the proton radius
of $^{56}$Ni are extrapolated values.\label{tab:one}}
\end{center}
\end{table}

\section{Results and discussion}

\label{sec:res}

\begin{figure}[htbp]
\begin{center}
\includegraphics[height=0.75\linewidth,angle=270]{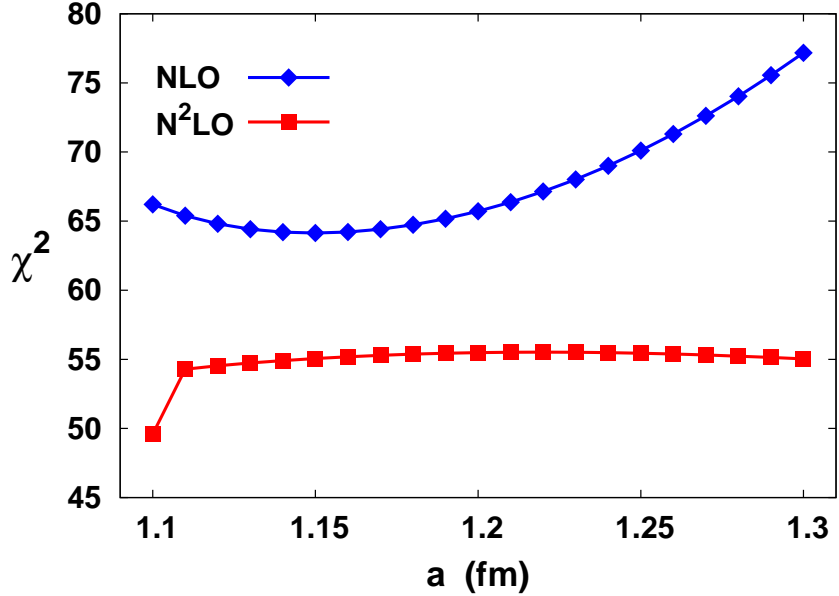}
\caption{(Color online) The NLO and N$^2$LO penalty functions as functions of the
regularization range $a$.\label{fig:chi2}}
\end{center}
\end{figure}

For the regularization ranges fixed at values between $a=1.1$ and
1.3\,fm, we minimised the penalty function defined in
section~\ref{sec:penalty}. At NLO and N$^2$LO, this corresponds to a
minimisation in 9- and 13-dimensional parameter space, respectively.
In Fig.~\ref{fig:chi2}, we show the obtained values of the NLO and
N$^2$LO penalty functions (\ref{chi3}) as functions of $a$. As one
could expect, the N$^2$LO penalty function is always lower than that
at NLO. If we split different contributions to the penalty function,
as defined in Eq.~(\ref{chi3}), we observe that the main improvement comes
from the properties of the saturation point and from the gap in
$^{120}$Sn, see Table~\ref{tab:contrib}.

\begin{table}[htbp]
\caption[T]{Contributions to the penalty function,
as defined in Eq.~(\protect\ref{chi3}), along with its total value,
shown at the regularization range of $a=1.15$\,fm.\label{tab:contrib}}
\vspace*{0.5cm}\hspace*{2.6cm}\begin{tabular}{|l|rrrrrr|r|}
\hline
        & \multicolumn{1}{c}{$\chi^2_\mathrm{inm}$}
        & \multicolumn{1}{c}{$\chi^2_\mathrm{pol}$}
        & \multicolumn{1}{c}{$\chi^2_\mathrm{BE}$}
        & \multicolumn{1}{c}{$\chi^2_\mathrm{rad}$}
        & \multicolumn{1}{c}{$\chi^2_\mathrm{gap}$}
        & \multicolumn{1}{c|}{$\chi^2_{\rho_1}$}
        & \multicolumn{1}{c|}{$\chi^2$} \\
\hline
NLO     & 14.413 & 0.158 & 43.752 & 0.905 & 3.757 & 1.153 & 64.138 \\
\hline
N$^2$LO &  5.374 & 0.134 & 44.491 & 2.884 & 1.840 & 0.336 & 55.059 \\
\hline
\end{tabular}
\end{table}

In the interval 1.1--1.3\,fm, the NLO penalty function shows a
significant dependence on the regularization range, whereas at
N$^2$LO it is almost flat. This means that already at N$^2$LO, a
change of the regularization range can be absorbed in the
readjustment of coupling constants, as prescribed by the
effective-theory approach~\cite{(Dob12b)}.

At NLO, we were able to continue the minimisation to values of $a$
well outside the interval 1.1--1.3\,fm. However, at N$^2$LO, near
$a=1.1$\,fm, we observe a sudden decrease of the penalty function. This
decrease is accompanied by a strong readjustment of the coupling
constants and a dramatic increase of the number of iterations needed
to converge the HFB iterations for all nuclei. This trend becomes
more pronounced for $a<1.1$\,fm, and sometimes leads to
impossibility to converge some of the HFB calculations defining the
penalty function. We tentatively attribute these feature to the
development of finite-size instabilities, which were not kept under
control by the constraint on the isovector density in $^{208}$Pb. We
decided to leave any further investigation of this feature till when
a linear response code is developed for finite-range EDF generators.
Consequently, we did not continue to adjust the N$^2$LO
pseudopotentials for shorter regularization ranges.
We note here, that the appearance of instabilities at short
regularization ranges may simply be the result of the parameter space
being restricted by conditions (\ref{eq:local}), that is, by using
local generators. Indeed, owing to these conditions, the Skyrme
generators cannot be obtained by bringing to zero the regularization
ranges of local generators (\ref{eq:NLOpot1}) or (\ref{eq:NLOpot2}).
Therefore, studies in this limit are deferred till when restrictions
to local generators are released.

\begin{table}[htbp]
\caption[T]{The NLO and N$^2$LO coupling constants of  local pseudopotentials
(\protect\ref{eq:LOpot}) and (\protect\ref{eq:locpot}) regularized
at $a=1.15$\,fm.\label{tab:CC} (in MeV\,fm$^{n+3}$) shown together with
their statistical errors.}
\vspace*{0.5cm}\hspace*{2.6cm}\begin{tabular}{|l|c|rr|}
\hline
          \multicolumn{1}{|c|}{Order}
        & \multicolumn{1}{c|}{Coupling}
        & \multicolumn{1}{c}{NLO}
        & \multicolumn{1}{c|}{N$^2$LO} \\
        & \multicolumn{1}{c|}{Constant}
        & \multicolumn{1}{c}{REG2c.161026}
        & \multicolumn{1}{c|}{REG4c.161026} \\
\hline
$n=0$   &$W_1^{(0)}$     &     41.678375$\pm$0.6 &      3121.637124$\pm$1.5 \\
        &$B_1^{(0)}$     &$-$1405.790048$\pm$4.3 &   $-$4884.029523$\pm$1.8 \\
        &$H_1^{(0)}$     &    202.879894$\pm$4.1 &      3688.310059$\pm$2.9 \\
        &$M_1^{(0)}$     &$-$2460.684507$\pm$6.7 &   $-$5661.028710$\pm$2.8 \\
         \hline
$n=2$   &$W_1^{(2)}$     &  $-$79.747992$\pm$4.2 &       547.802973$\pm$1.9 \\
        &$B_1^{(2)}$     &     73.112729$\pm$1.4 &    $-$319.513120$\pm$1.3 \\
        &$H_1^{(2)}$     & $-$681.295790$\pm$3.2 &    $-$134.164127$\pm$0.3 \\
        &$M_1^{(2)}$     &  $-$48.161707$\pm$5.1 &    $-$318.407541$\pm$0.6 \\
         \hline
$n=4$   &$W_1^{(4)}$     &                       &      2019.945667$\pm$2.2 \\
        &$B_1^{(4)}$     &                       &   $-$2365.956384$\pm$1.6 \\
        &$H_1^{(4)}$     &                       &      2310.445509$\pm$1.8 \\
        &$M_1^{(4)}$     &                       &   $-$2117.509518$\pm$4.0 \\
         \hline
        &$W_\mathrm{SO}$ &    177.076480$\pm$4.7 &       174.786236$\pm$5.1 \\
\hline
\end{tabular}
\end{table}

The NLO and N$^2$LO coupling constants as functions of the
regularization range $a$ are plotted in the supplemental
material~[URL]. In the rest of this article, we
discuss results obtained for $a=1.15$\,fm, which approximately
corresponds to the minimum of the penalty function for the
pseudopotential at NLO. Numerical unrounded values of the
coupling constants at $a=1.15$\,fm are listed in Table~\ref{tab:CC}.
Following the naming convention introduced
in~\cite{Bennaceur:14}, we call parameter sets of regularized
potentials as REGnx.DATE, where $n=2p$ is the maximum order of
higher-order differential operators used at N$^p$LO, letter ``x''
distinguishes different versions of the implementation, and ``DATE''
is a time stamp. In this study, we put x$\rightarrow$c to mark the fact that the
regularized potential is local, it is accompanied by zero-range two-body
central and spin-orbit forces, and it is evaluated along with
two-body centre-of-mass correction and exact direct and exchange
Coulomb terms.

\subsection{Infinite nuclear matter}

\begin{figure}[htbp]
\begin{center}
\includegraphics[height=0.8\linewidth,angle=270]{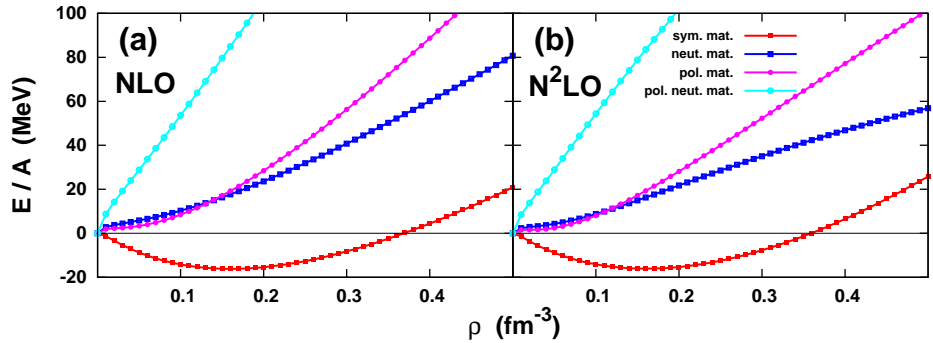}
\caption{(Color online) Infinite-nuclear-matter equations of state for the NLO and N$^2$LO pseudopotentials
at the regularization range of $a=1.15$\,fm.\label{fig:eos1}}
\end{center}
\end{figure}

\begin{figure}[htbp]
\begin{center}
\includegraphics[height=0.8\linewidth,angle=270]{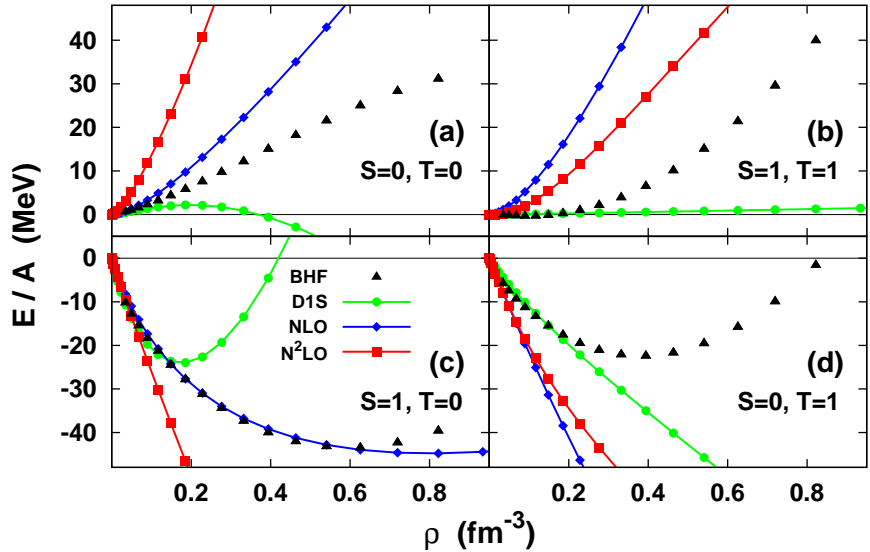}
\caption{(Color online) Infinite-symmetric-nuclear-matter equation of state
calculated for the NLO and N$^2$LO pseudopotentials at the
regularization range of $a=1.15$\,fm and split into four
(iso)scalar/(iso)vector channels. Results are compared with those obtained
for the Gogny D1S force~\cite{BERGER1991365} and
Brueckner-Hartree-Fock (BHF) calculations of~\cite{(Bal97),BaldoPrivate}.
\label{fig:est}}

\end{center}
\end{figure}

Figure~\ref{fig:eos1} shows the energy per nucleon for different
states of infinite nuclear matter obtained with the pseudopotentials
at NLO and N$^2$LO regularized at $a=1.15$\,fm. Near the saturation
point, the two resulting equations of state are qualitatively
similar. Nonetheless, for pure neutron matter and polarized symmetric
matter, one can observe a trend that is significantly different for
the high density part of the equations of state, where at N$^2$LO the energy per
particle increases less rapidly. This high density region was not
constrained, and it is not expected to have a sizable impact on
properties of finite nuclei. Similarities between the NLO and N$^2$LO
equations of state partly come from strong constraints that were put
on the properties of the saturation point. These properties are
indeed well reproduced by the two pseudopotentials, see
Table~\ref{tab:sat}.

\begin{table}[htbp]
\caption{Properties of the saturation point of infinite nuclear matter
obtained for the NLO and N$^2$LO pseudopotentials regularized at $a=1.15$\,fm.
\label{tab:sat}}
\vspace*{0.5cm}\hspace*{1.6cm}\begin{tabular}{|l|cccccc|}
\hline
 & $\rho_\mathrm{sat}~(\mathrm{fm}^{-3})$ & $B$~(MeV) & $K_\infty$~(MeV)
   & $m^*/m$ & $J$~(MeV) & $L$~(MeV) \\
\hline
NLO     & 0.1599 & -16.17 & 229.8 & 0.4076 & 31.96 & 64.04 \\
\hline
N$^2$LO & 0.1601 & -16.09 & 230.0 & 0.4061 & 31.95 & 64.68 \\
\hline
\end{tabular}
\end{table}

In Fig.~\ref{fig:est}, we show the NLO and N$^2$LO equation of state
of infinite symmetric nuclear matter split into four
(iso)scalar/(iso)vector channels. As discussed
in~\cite{Davesne:15}, in principle the N$^2$LO pseudopotential can
be accurately adjusted to reproduce all four channels simultaneously.
In our case, because of the low effective mass, we had to overshoot
the strength of the interaction in the scalar-isovector ($S=0, T=1$)
channel, Fig.~\ref{fig:est}(d). As the sum of all four channels gives
the constrained symmetric nuclear matter, equations of states in the
$S=0, T=0$ and $S=1, T=1$ channels came out too high. Similarly as
in~\cite{Davesne:15}, where an effective density-dependent term has
been added on top of the N$^2$LO pseudopotential, we can expect that this unwelcome feature can
be corrected in future implementations involving a three-body force.

\subsection{Statistical error analysis}
\label{error}

\begin{figure}[htbp]
\begin{center}
\includegraphics[height=0.35\linewidth]{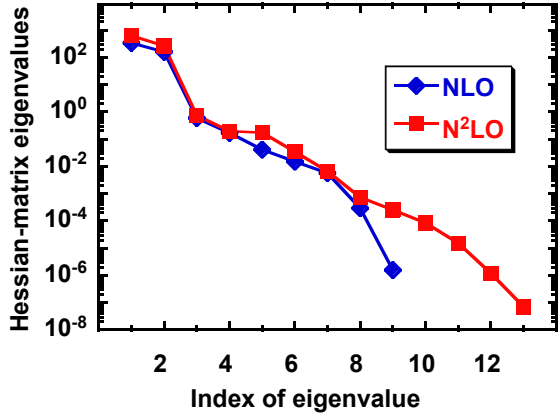}
\caption{(Color online) Eigenvalues of the Hessian matrices
calculated from the normalized penalty function for pseudopotentials
at NLO and N$^2$LO with the regularization range of $a=1.15$\,fm.
\label{fig:reg-eigenvalue}}
\end{center}
\end{figure}

For the two pseudopotentials built at NLO and N$^2$LO with
regularization range $a=1.15$\,fm, we performed the analysis of
statistical errors along the lines presented
in~\cite{0954-3899-41-7-074001}. For that purpose, we considered
the scaled penalty function $\chi^2_\mathrm{norm}$, for which we
calculated the Hessian matrix. Its eigenvalues are shown in
Fig.~\ref{fig:reg-eigenvalue}. The total number of eigenvalues corresponds
to the number of parameters allowed to vary during the fit, that is,
to 9 for the pseudopotential at NLO and to 13 at N$^2$LO.

Eigenvalues of the Hessian matrix are indicative of how well the
penalty function is constrained in those directions of the parameter
space that are given by its eigenvectors. From the gap between the
second and third eigenvalue it clearly appears that, irrespective of
the order at which the pseudopotential is built, two such directions
are well constrained. Beyond this second eigenvalue, the eigenvalues
decrease in a fairly regular manner, and it is not possible to
unambiguously define a dividing point between relevant and irrelevant
eigenvalues. Furthermore, we have checked that the directions given
by the eigenvectors of the Hessian matrix mix all the terms of the
pseudopotential, so that no coupling constant (for the
parametrization we have adopted) can be removed or frozen to get rid
of a specific small eigenvalue.

\begin{figure}[htbp]
\begin{center}
\includegraphics[height=0.80\linewidth,angle=270]{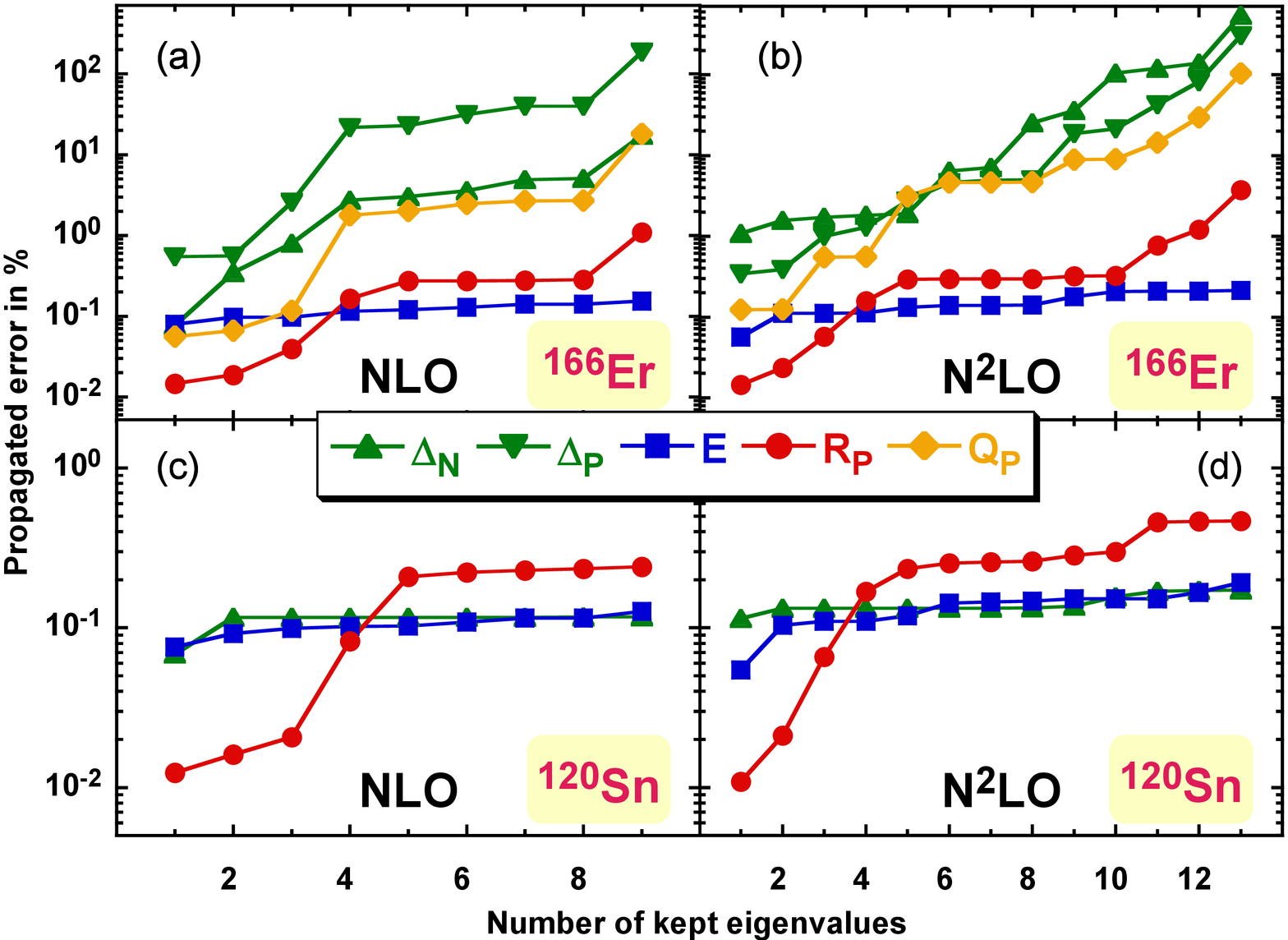}
\caption{(Color online) Lower panels: Propagated errors of the total
binding energies ($E$), average neutron pairing gaps ($\Delta_N$),
and proton rms radii ($R_P$) calculated for the spherical nucleus
$^{120}$Sn as a functions of the number of smallest eigenvalues kept
in the spectrum of the covariance matrix. Upper panels: the same
quantities supplemented with the proton average pairing gaps
($\Delta_P$) and proton quadrupole moments ($Q_P$) calculated for the
deformed nucleus $^{166}$Er. All values are in per cent.
\label{fig:reg-propagated}}
\end{center}
\end{figure}

The covariance matrix is the inverse of the Hessian matrix with a
given number of its eigenvalues kept~\cite{0954-3899-41-7-074001}.
Its average value, calculated for a vector of derivatives of a given
observable with respect to the parameters of the model, is called propagated
error of the observable. In Fig.~\ref{fig:reg-propagated}, we show
such propagated errors determined for several observables in
$^{120}$Sn and $^{166}$Er as functions of the number of largest
eigenvalues kept in the spectrum of the Hessian matrix.

We see that for the two considered nuclei, the propagated errors are
qualitatively similar when going from NLO to N$^2$LO. Errors of
the total binding energies are generally small, of the order of
0.1~\%, and do not show any significant dependence on the number of
kept eigenvalues. For $^{120}$Sn, the errors of the average neutron
pairing gaps are similarly small and almost flat. This means that
these observables, which were included in the definition of the
penalty function, section~\ref{sec:penalty}, are insensitive to the
unconstrained directions in the parameter space. Similarly, flatness
of the propagated errors of the binding energy of $^{166}$Er, which
was not included in the penalty function, indicates that adding this
one observable to the penalty function would not help in better
constraining the model parameters.

The propagated errors of the proton rms radii are only about 0.01~\%
in the extreme case when only one eigenvalue is kept, and for five or
more eigenvalues they reach what is approximately a plateau of
0.2--0.4~\%. This means that directions in the parameter space that
are associated with about five largest eigenvalues are meaningful, and
should be included in the covariance matrix. Moreover, for
$^{166}$Er, the propagated errors of radii increase significantly when
the last eigenvalue (NLO), or three last eigenvalues (N$^2$LO), are
taken into account. This means that the current penalty function, which does not
carry information concerning the structural properties of
deformed nuclei, would be enriched by incorporating such information
in future parameter adjustments. This conclusion is substantiated by
the propagated errors of quadrupole moments, which increase from
0.1~\% to even 100~\%.

The main, and striking, difference between the results obtained for
$^{120}$Sn and $^{166}$Er concerns the average pairing gaps. As we
already said, in $^{120}$Sn, the propagated errors of the neutron gap
are small and almost flat. However, in $^{166}$Er, one sees a clear
gradual increase of errors of pairing gaps with the number of kept
eigenvalues. These large propagated errors are likely associated with
the sensitivity of gaps to deformation. Indeed, changes of coupling
constants induce changes of shape of the $^{166}$Er ground state, and
thus changes of its single-particle spectrum. This, in turn, can
significantly modify the average pairing gaps and lead to large
calculated propagated errors. Altogether, we conclude that adding to
the penalty function data on spectroscopic properties of deformed
nuclei may be more interesting from the point of view of constraining
the model parameters than adding those related to their bulk properties
like masses or radii.

For the following part of this article, to calculate the propagated
errors, we chose to keep five largest eigenvalues of the Hessian
matrices. This choice is based on the observation that beyond this point, several
propagated errors, like those of radii, stop changing in a significant way.
The corresponding covariance matrices are provided in the
supplemental material~[URL]. In Table~\ref{tab:CC} we show statistical
errors of the coupling constants, which are equal to square roots
of diagonal elements of the covariance matrices~\cite{0954-3899-41-7-074001}.
Note that in Table~\ref{tab:CC} we show unrounded values of the
coupling constants, with several more digits beyond the statistically
significant ones. Nevertheless, performing calculations with properly
rounded values does change results, and significantly increases
values of the penalty functions that move away from their minima.
These changes are, of course, within bounds of propagated errors and
thus are statistically insignificant, however, they also spoil smooth
behaviour of observables and parameters as functions of the
regularization range $a$. Therefore, in all calculations we recommend
using the unrounded values of the coupling constants. Note also, that
errors of parameters serve to illustrate the overall uncertainty of
parameters only, whereas proper propagated errors of observables must
be obtained by using the full covariance matrices.

\subsection{Finite nuclei}

\begin{figure}[htbp]
\begin{center}
\includegraphics[height=0.5\linewidth,angle=270]{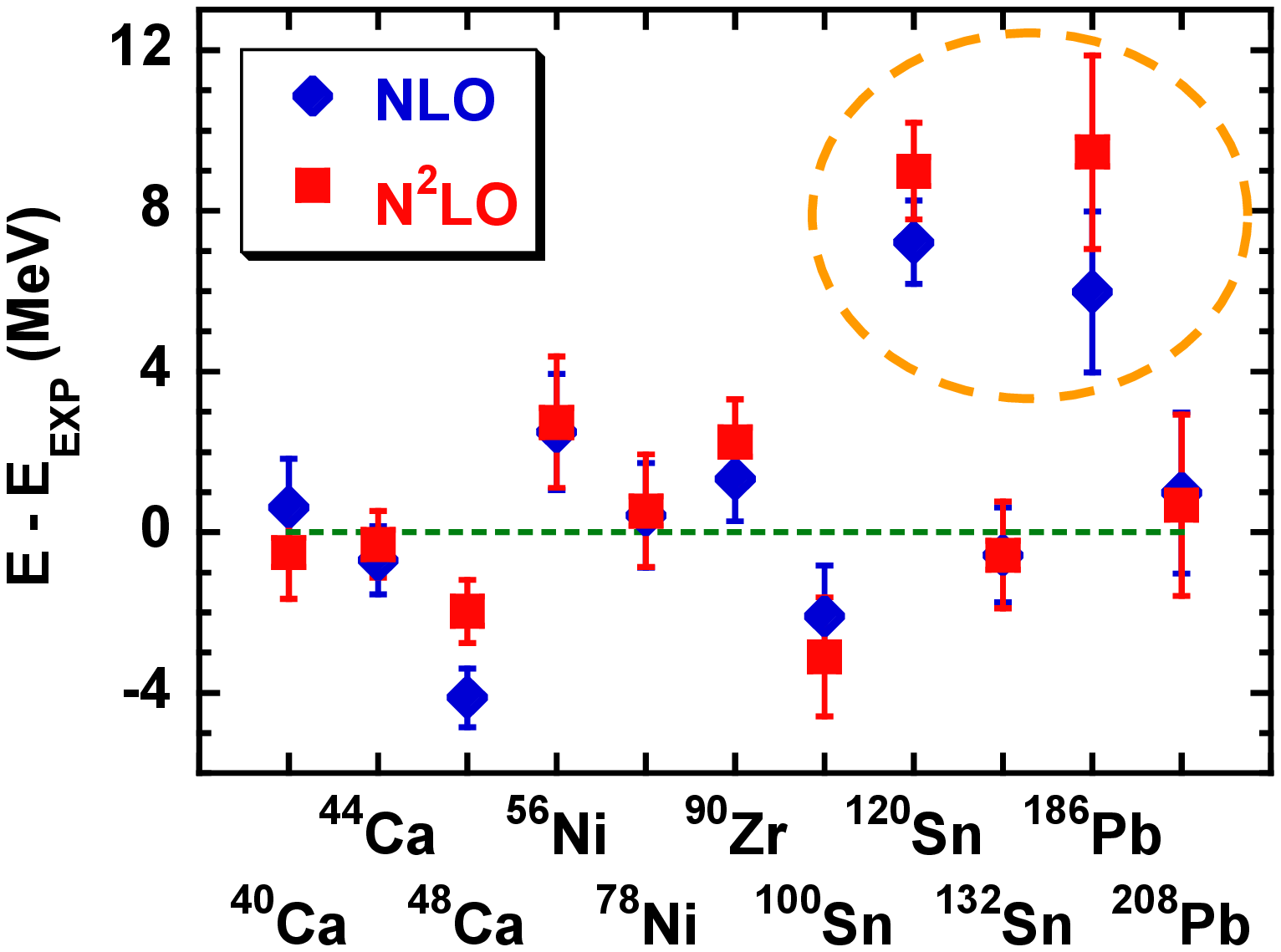}
\caption{(Color online) Ground-state energies of selected spherical
nuclei and their propagated errors calculated for the NLO and N$^2$LO
pseudopotentials regularized at $a=1.15$\,fm, relative to experiment.
The two open-shell outliers $^{120}$Sn and $^{186}$Pb (discussed in the text)
are highlighted by the orange ellipse.
\label{fig:reg-exp}}
\end{center}
\end{figure}

\begin{figure}[htbp]
\begin{center}
\includegraphics[height=0.80\linewidth,angle=270]{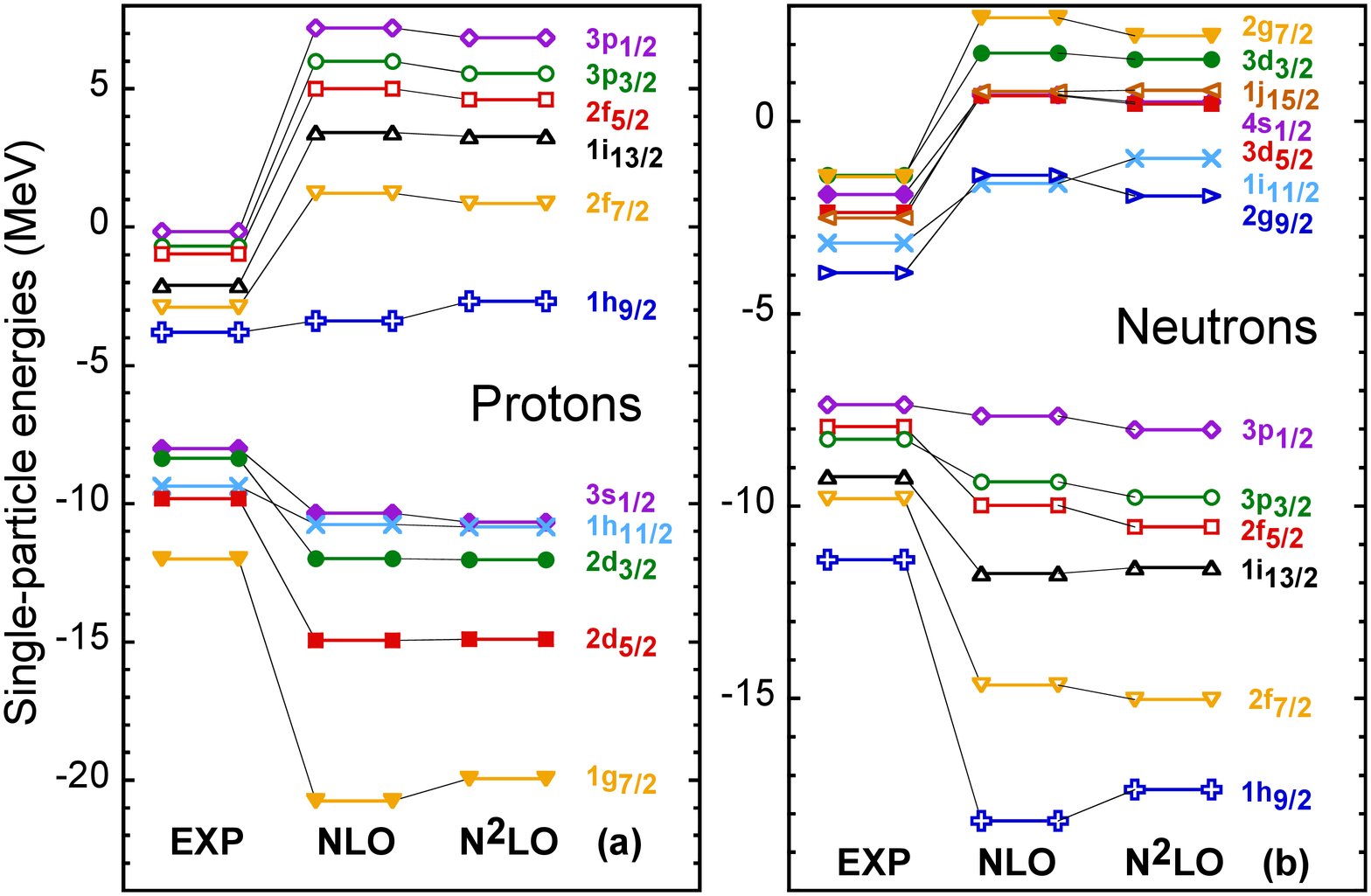}
\caption{(Color online) Proton (left) and neutron (right) single-particle
energies calculated in $^{208}$Pb in comparison with the empirical
values taken from Ref.~\protect\cite{[Sch07c]} Note that states appearing
at positive energies rather correspond to single-particle resonanses
estimated by using the finite harmonic-oscillator basis.\label{Pbsp}}
\end{center}
\end{figure}

In Fig.~\ref{fig:reg-exp}, we show ground-state energies of selected
spherical nuclei obtained for the NLO and N$^2$LO pseudopotentials
regularized at $a=1.15$\,fm. In addition to nuclei that were used to
build the penalty function, results are shown for $^{44}$Ca,
$^{90}$Zr, and $^{186}$Pb.
Apart from $^{48}$Ca, $^{120}$Sn, and $^{186}$Pb, the agreement of
the calculated binding energies with the experimental data is
compatible with the calculated propagated errors. Large deviations
obtained for two outliers, $^{120}$Sn and $^{186}$Pb, can most
probably be related to the low effective mass and the resulting
unrealistically small density of single-particle states. This is
illustrated in Fig.~\ref{Pbsp}, where we show proton and neutron single-particle
energies calculated in $^{208}$Pb in comparison with the empirical
values taken from Ref.~\cite{[Sch07c]}.

As can be seen in Figs.~\ref{fig:reg-exp} and~\ref{Pbsp}, results
obtained for both pseudopotentials are fairly similar, and we do not
see any significant improvement when going from NLO to N$^2$LO. This
is also visible in Table~\ref{tab:contrib}, where the decrease of the
penalty function is mostly related to the improvement of
nuclear-matter properties.

\begin{figure}[htbp]
\begin{center}
\includegraphics[height=0.35\linewidth]{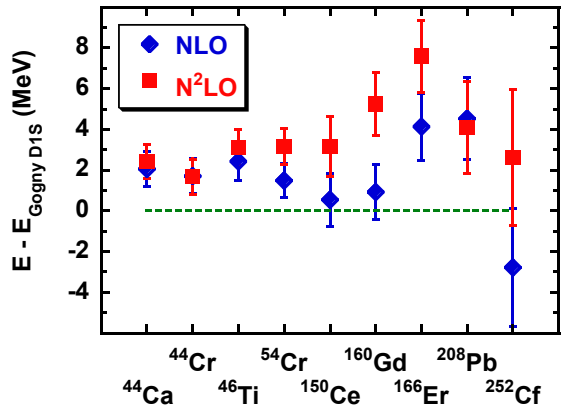}
\caption{(Color online) Ground-state energies of selected spherical
and deformed nuclei and their propagated errors calculated for the
NLO and N$^2$LO pseudopotentials regularized at $a=1.15$\,fm, relative
to those obtained for the Gogny D1S
force~\cite{BERGER1991365}.\label{fig:reg-energy}}
\end{center}
\end{figure}

Since our long term goal is to develop a pseudopotential for beyond mean-field
calculation, a direct comparison of calculated mean-field binding energies
with experimental data, as we did with Figure~\ref{fig:reg-exp},
only gives a partial information. It is of interest
to compare the mean-field results with other calculations done at
the same approximation from an effective interaction which is routinely
used for mean-field calculations~\cite{HFBD1S} and some beyond mean-field ones.

In Fig.~\ref{fig:reg-energy} we compare ground-state energies
calculated for the NLO and N$^2$LO pseudopotentials with those
obtained for the Gogny D1S force~\cite{BERGER1991365}. These
calculations were performed using finite harmonic-oscillator basis.
Therefore, the results carry some trivial offset with respect to how the
parameters were adjusted; nevertheless, the same basis was used for
all three sets of calculations, and thus the relative energies are fully
meaningful. We see that extrapolations to deformed nuclei, which were
not included in the penalty function, work fairly well, although
the propagated errors clearly increase with mass.

\begin{figure}[htbp]
\begin{center}
\includegraphics[height=0.35\linewidth]{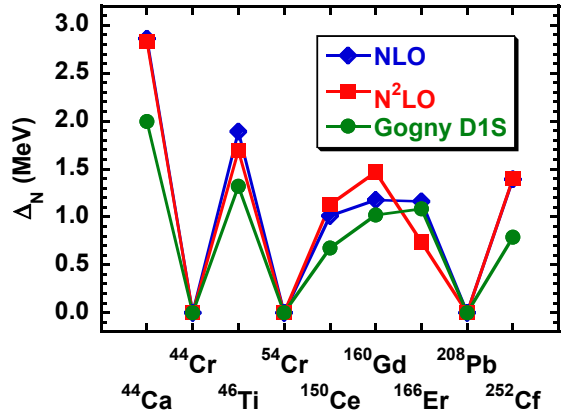}
\caption{(Color online) Average neutron pairing gaps of selected spherical and
deformed nuclei (their propagated errors are smaller than the symbols)
calculated for the NLO and
N$^2$LO pseudopotentials regularized at $a=1.15$\,fm, compared with those
obtained for the Gogny D1S force~\cite{BERGER1991365}.
\label{fig:reg-deltaN}}
\end{center}
\end{figure}

Finally, in Fig.~\ref{fig:reg-deltaN}, we plot the average neutron
gaps calculated for the NLO and N$^2$LO pseudopotentials and the
Gogny D1S force~\cite{BERGER1991365}. As anticipated by the
constraint that we used for the neutron gap in $^{120}$Sn in the
penalty function, neutron gaps obtained for the pseudopotentials are
fairly large, and almost for all nuclei considered here exceed those
given by D1S. The primary goal of this preliminary study was to check
if the form of the pseudopotential we develop allows us to generate
interaction that is enough attractive in the particle-particle
channel to lead to significant gaps. Fig.~\ref{fig:reg-deltaN} shows
that this goal was achieved. However, we also obtained proton gaps
that often collapse in the deformed minima. This latter unwanted feature may
once again be due to the low effective mass and too small density of
single-particle energies.

\section{Conclusions}
\label{sec:Conclusions}

In this article, we considered a local finite-range pseudopotential
with a Gaussian regulator and derived its contributions to the
particle-hole and particle-particle energies and to mean-field
equations up to N$^2$LO. When supplemented with a zero-range two-body
force acting in the particle-hole channel, it allowed us to build the
first spuriousity-free nonlocal energy density functional that is
capable of describing paired finite nuclei without using
density-dependent terms.

To adjust the coupling constants of the pseudopotential, we defined a
penalty function that constrained properties of selected spherical
nuclei along with those of infinite nuclear matter. Its behaviour
near the minimum allowed us to evaluate statistical errors of
coupling constants and propagated errors of observables, including
those for selected deformed nuclei. In this way, we were able to
determine well and poorly constrained directions in the parameters
space.

We consider the present parameterization to be an initial step
towards a more definite solutions only. At present, the
pseudopotential considered here gives low values of the effective
mass, and consequently low densities of single-particle states in
finite nuclei. In future implementations, we plan to correct for this
by introducing three-body (or four-body) terms, eventually
supplemented by velocity-dependent terms~\cite{(Sad13d)}. Our
approach would then correspond to an extension of the interaction
proposed by Onishi and Negele~\cite{ONISHI1978336}, who showed that
it may have encouraging results for mean-field calculations.

The identified deficiencies of the NLO and N$^2$LO pseudopotentials
derived in this work preclude using it within massive calculations
that would produce masses and collective spectra across the Segr\'e
chart. We will undertake this task once we correct for this
deficiencies, as mentioned above. Nevertheless, in Ref.~\cite{[Ben17b]} we
present and discuss results of calculations performed in semi-magic
nuclei, and we refer the Reader to this publication for further
information.

\ack{We wish to thank Marcello Baldo for providing us with his results from
Brueckner-Hartree-Fock calculations of infinite nuclear matter and Michael
Bender for a careful reading of this manuscript and helpful comments.
K.B.\ thanks Tom\'as Rodriguez and Nicolas Schunck for useful discussions and
benchmark calculations during the development of the code {\sc{finres}}$_4$.
This work was supported by the Academy of Finland and University of
Jyv\"askyl\"a within the FIDIPRO program, by the Royal Society and
Newton Fund under the Newton International Fellowship scheme, by the
CNRS/IN2P3 through PICS No.\ 6949, by the Polish National Science
Center under Contract No.\ 2012/07/B/ST2/03907, and by the Academy of
Finland under the Centre of Excellence Program 2012–2017 (Nuclear and
Accelerator-Based Physics Program at JYFL). We acknowledge the CSC-IT
Center for Science Ltd., Finland, for the allocation of computational
resources.}

\bigskip

\section*{References}

\bibliographystyle{iopart-num}

\providecommand{\newblock}{}

\end{document}